\documentclass[aps,preprint,eqsecnum,floatfix]{revtex4}
\usepackage{graphics}
\begin{document}
\title{Parity-Violating Interactions and Currents in the Deuteron}
\author{R.\ Schiavilla}
\affiliation{Jefferson Laboratory\\ Newport News, VA 23606\\
and\\
Physics Department\\ Old Dominion University \\ Norfolk, VA 23529}
\author{J.\ Carlson and M.\ Paris}
\affiliation{Theoretical Division\\ Los Alamos National Laboratory\\ Los Alamos, NM 87545}
%
%
%
%
\begin{abstract}
We investigate parity-violating asymmetries in $\vec{n}p$ radiative
capture at thermal neutron energies and in deuteron electro-disintegration
in quasi-elastic kinematics, using the DDH model for the parity-violating
nucleon-nucleon interaction.  We find dramatic cancellations between the
asymmetries induced by the parity-violating interaction and those arising
from the associated parity-violating pion-exchange currents.  In the
$\vec{n}p$ capture, the model-dependence of the result is nevertheless
quite small, because of constraints arising through the Siegert evaluation
of the relevant $E_1$ matrix elements.  In quasi-elastic electron scattering
these processes are found to be insignificant compared to the asymmetry
produced by $\gamma$-$Z$ interference on individual nucleons.  These two
experiments, then, provide clean probes of different aspects of weak-interaction
physics associated with parity violation in the $np$ system.
\end{abstract}
%
%
\maketitle
\section{Introduction}

Two-nucleon experiments are the clearest probes of hadronic parity violation.
A recent experiment has measured the longitudinal asymmetry in $\vec{p}p$
elastic scattering~\cite{Berdoz01}, while another one is underway to measure
the photon asymmetry in low-energy $\vec{n}p$ radiative capture~\cite{npexpt}.
This experiment has been designed to provide a definitive measurement of the
weak $\pi$$N$$N$ coupling constant, which determines the longest-range part of
the parity-violating nucleon-nucleon interaction. 

We investigate this process using the Desplanques, Donoghue, and Holstein (DDH)
meson-exchange model of the parity-violating $N$$N$ interaction~\cite{ddhpv}.  
We separately evaluate the contributions from the hadronic weak interaction and
the associated two-body currents.  These currents play an important role, reducing
dramatically the measured asymmetry.  We also consider the model-dependence of
the full result by using different modern models of the strong interaction.  Even
though there are significant cancellations between the two terms above, the final
model dependence is quite small, and in fact similar in size to the estimated
contribution of short-range mechanisms.

It has been speculated that these interactions and currents could potentially 
also play 
a role in the parity-violating quasi-elastic electro-disintegration of the 
deuteron, recently measured in the SAMPLE experiments at the MIT-Bates
facility~\cite{Hasty00,samplereview}.  Indeed, their contributions could cloud the
interpretation of the experimental asymmetry in terms of single-nucleon properties.
However, we find that these two-nucleon mechanisms lead, for any reasonable value
of the $\pi$$N$$N$ coupling constant, to very small asymmetries when compared to
those originating from $\gamma$-$Z$ interference.  Therefore, these electron scattering
results can be reliably used to extract single-nucleon properties.

A more complete account of the calculations carried out so far will be published
elsewhere~\cite{Carlson02}.  In the present letter, we only report the main results.

\section{Asymmetry in $\vec{n}p$ radiative capture}

The parity-violating (PV) observable in the $\vec{n}p$ radiative capture--the asymmetry
in the angular distribution of the outgoing photon with respect to the direction of the
initial neutron polarization--is very sensitive to the weak parity-violating $\pi$$N$$N$
coupling~\cite{npdgtheory1,npdgtheory2}.  Since this coupling contributes to the
longest-range part of the interaction, it provides a window into the hadronic weak interaction
in a manner similar to the role the standard $\pi$$N$$N$ coupling provides in strong-interaction
physics.  Its experimental determination, however, has proven to be very difficult.  Measurements
of the circular polarization of gamma-ray decays in $^{18}$F~\cite{flourine} have been
interpreted to indicate a very small $\pi$$N$$N$ coupling as compared to theoretical expectations
based upon hadronic models~\cite{ddhpv}, while measurements in atomic $^{133}$Cs seem to favor
a much larger value~\cite{cesium}.  The analysis of both these experiments is complicated
by the difficulty of handling reliably, from a theoretical stand-point, the many-body aspects of
the problem in these complex systems.  These issues have been addressed in Refs. \cite{haxtonreview} and \cite{Haxton02}.
More recently, PV asymmetry measurements of neutron resonances in compound nuclei seem to
require even larger values of the weak $\pi$$N$$N$ coupling constant~\cite{bowmanprivate}.

For these reasons, an experiment measuring the photon asymmetry in $\vec{n}p$ capture has
been undertaken at the LANSCE facility~\cite{npexpt}.  As mentioned above, previous studies
of this process~\cite{npdgtheory1,npdgtheory2} have shown that this asymmetry is very sensitive
to the weak $\pi$$N$$N$ coupling constant, while essentially unaffected by short-range
contributions.  Here we investigate the model dependence of this result by considering
several different high-quality interactions fit to the strong-interaction data.  In addition,
we consider the individual contributions to the final result, including processes
where the photon couples at all points to the exchanged (virtual) pion.

We employ the standard DDH meson-exchange model of the (PV) $N$$N$ interaction, solving
the full Schr\"odinger equation for the scattering state as well as the deuteron bound
state.  The equations and relevant mixing parameters in the $S$-matrix have been discussed
in considerable detail in Ref.~\cite{pppvcalc} for the case of $\vec{p}p$ elastic scattering.
While the short-range contributions to the PV interaction should not be viewed as resulting
only from the exchange of single mesons, the seven parameters of the DDH model can still
be employed to characterize all the low-energy PV mixings.  For example, two-pion exchange
could play a role~\cite{Pirner73}, however we assume that its effects can be included, at
least at low energy, through the present combination of pion- and short-range terms.  The
complete expressions for the $S$-matrix and contributions of the various terms will be
presented in Ref.~\cite{Carlson02}.

We calculate the asymmetry for the AV18~\cite{av18}, Bonn-CD~\cite{bonncd} and
Nijmegen-I (NIJM-I)~\cite{nijm1} interactions.  Each strong interaction model has associated
exchange currents.  For the AV18 we include currents from the momentum-independent
terms ($\pi$- and $\rho$-currents) as well as terms from the momentum-dependent
terms in the interaction.  Further discussion of the AV18 currents
is given below, for a review see Ref.~\cite{pccurrents}.  For the Bonn-CD
and NIJM-I interactions, we include $\pi$- and $\rho$-currents with the cutoffs from
the Bonn-CD model ($\Lambda_\pi$=1.72 GeV and $\Lambda_\rho$=1.31 GeV).  Contributions
from other meson exchanges in the Bonn and Nijmegen models have been neglected.  In all
calculations, the currents associated with $\Delta$ excitation and $\omega \pi \gamma$
transition have been included~\cite{pccurrents}.

The total cross section at thermal neutron energies is due to the well-known $M_1$
transition connecting the parity-conserving (PC) $^1$S$_0$ $n$$p$ state to the
PC deuteron state.  The calculated values for each model are given in Table~I,
both for one-body (impulse) currents alone and for the one- and two-body
currents.  In each case the largest two-body contribution, approximately
two-thirds of the total, comes from the currents associated with pion
exchange.  The total cross section is in good agreement with experimental
results, which are variously quoted as 334.2(0.5) mb~\cite{npxsec1} or
332.6(0.7) mb~\cite{npxsec2}.  It would be possible to adjust, for example,
the transition magnetic moment $\mu_{\gamma N\Delta}$ 
of the $\Delta$-excitation current to precisely fit one of these
values, here we simply choose a $\mu_{\gamma N\Delta}$ of 3 n.m.,
which is consistent with an analysis of $\gamma$-$N$ data at resonance.

The PV asymmetry arises from an interference between the $M_1$ term above 
and the $E_1$ transition, connecting the $^3$P$_1$ PV $n$$p$ state
to the PC deuteron state and the $^3$S$_1$ PC $n$$p$ state to the $^3$P$_1$
PV deuteron state.  The PV components of the wave functions are generated by
the DDH potential, including $\pi$-, $\rho$-, and $\omega$-exchange mechanisms.
In this work we have taken the linear combination of $\rho$- and $\omega$-weak
coupling constants corresponding to $p$$p$ scattering from an earlier
analysis~\cite{pppvcalc} of these experiments.  The remaining couplings have
been taken from the DDH \lq\lq best guess\rq\rq estimates.  Finally, as in the
earlier analysis of $\vec{p}p$ scattering, the cutoff values in the meson-exchange
interaction are taken from the Bonn-CD potential ($\Lambda_\pi$=1.72 GeV,
$\Lambda_\rho$=1.31 GeV, and $\Lambda_\omega$=1.50 GeV).

With these couplings, we obtain the asymmetries also listed in Table~I.
The results are consistent with earlier~\cite{npdgtheory1} and
more recent~\cite{Desplanques01} estimates, and all agree within
a few per cent, which is also the magnitude of the contributions from the
short-range terms.  The $E_1$ transition has been calculated in the long-wavelength
approximation (LWA) with the Siegert form of the $E_1$ operator, thus eliminating
many of the model dependencies and leaving only simple (long-range) matrix
elements.  We have explicitly calculated corrections beyond the LWA $E_1$ terms,
and found them to be quite small.

We have also calculated the $E_1$ contributions with the same current
operator used to calculate the $M_1$ matrix element.  To the extent
that retardation corrections beyond the LWA of the $E_1$ operator
are negligible~\cite{Viviani00}, this should produce identical results
{\it provided} the current is exactly conserved.  In order to satisfy current
conservation, currents from both the strong (PC) and weak (PV) interactions
are required.  In the following we keep only the $\pi$-exchange term in the
DDH interaction with their \lq\lq best guess\rq\rq~for the weak $\pi$-$N$
coupling constant, and use the AV18 strong-interaction model.

The PC $\pi$- and $\rho$-currents are derived from the $v_6$ part
of the AV18 interaction and by construction exactly satisfy current
conservation with it~\cite{Schiavilla89}.  The same
prescription is used to generate conserved PV $\pi$-currents from the
DDH interaction in the presence of a short-range cutoff (as is
the case here).  However, the PC currents originating from the momentum-dependent
terms of the AV18 are strictly not conserved.  For example, the currents
from the ${\bf L}^2$ and $({\bf L} \cdot {\bf S})^2$ components of the AV18 are 
constructed by minimal substitution~\cite{Schiavilla89}.  While this
procedure leads to conserved currents for the isospin-independent
${\bf L}^2$ and $({\bf L} \cdot {\bf S})^2$ terms, it is not
adequate for their isospin-dependent counter-parts, as one can
easily surmise by considering their commutator with the charge
density operator.  This commutator requires the presence of terms with
the isospin structure $({\tau}_i \times {\tau}_j)_z$, which
cannot be generated by minimal substitution (so-called internal
radiation contributions~\cite{Tsushima93}).  This issue will be discussed
more thoroughly in Ref.~\cite{Carlson02}.  

The currents associated with the momentum-dependent terms,
however, while small, play
here a crucial role, because of the large cancellation between
the (PC) $\pi$- and $\rho$-currents from the AV18 and the (PV)
$\pi$-currents from the DDH.  This point is illustrated in
Table~II.  Note that the PC currents from $\Delta$-excitation
and $\omega\pi\gamma$ transition mechanisms are transverse and
therefore do not affect the $E_1$ matrix element. However,
they slightly reduce the PV asymmetry since their
contributions increase the $M_1$ matrix element by $\simeq 1$\%.
They are not listed in Table~II.

The asymmetry is given by the sum of the two columns in Table~II,
namely $+0.17 \times 10^{-8}$ (last row).  This value should be
compared to $-5.02 \times 10^{-8}$, obtained with the Siegert
form of the $E_1$ operator for the same interactions (and currents
for the $M_1$ matrix element).  As already mentioned, we have explicitly
verified that retardation corrections in the $E_1$ operator are too small
to account for the difference.  Thus this difference is to be ascribed to the lack of
current conservation, originating from the isospin- and momentum-dependent
terms of the AV18.  

To substantiate this claim, we have carried out
a calculation based on a $v_6$ reduction~\cite{Wiringa02} of the
AV18, constrained to reproduce the binding energy of the deuteron
and some of the $N$$N$ phase shifts.  In this case, the
resulting PC currents (only the $\pi$- and $\rho$-exchange terms
are present) are exactly conserved.  The results are listed in
Table~III.  The remaning $\simeq 2.7$\% difference between the
Siegert result and the full calculation is presumably due
to numerical inaccuracies as well as additional corrections
from retardation corrections and higher order multipoles.
Both of these effects are included in the full calculation.  

\section{Asymmetry in $d(\vec{e},e^\prime)pn$ electro-disintegration}

The SAMPLE experiment\cite{Hasty00,samplereview} has measured the parity-violating
asymmetry in polarized electron quasi-elastic scattering on the
deuteron.  This asymmetry has two distinct contributions: one associated
with interference of the $\gamma$- and $Z$-exchange amplitudes, and
the other induced by PV $N$$N$ interactions.  The first contribution
was recently studied in Ref.~\cite{Diaconescu01}, where it was shown
that two-body terms in the nuclear electromagnetic and weak neutral
currents only produce 1--2\% corrections to the asymmetry due to the
corresponding single-nucleon currents.

In the present study, we investigate the asymmetry originating
from hadronic weak interactions.  We update and sharpen earlier
predictions obtained in Refs.~\cite{Hwang80,Hwang81}--these
studies did not include the effects of PV currents.

The expression for the asymmetry from gamma-Z interference
is given in Ref.~\cite{Diaconescu01}, while that due to the
hadronic weak interaction reads~\cite{Carlson02}

\begin{equation}
A(q,\omega,\theta_e)=\frac{v_T^\prime R^\prime_T(q,\omega)}
{v_L R_L(q,\omega)+v_T R_T(q,\omega)} \ ,
\end{equation}
where $v_L$, $v_T$, and $v_T^\prime$ denote electron kinematical
factors, $R_L$ and $R_T$ are the standard longitudinal and
transverse response functions (explicit expressions for these
and the $v$ factors are given in Ref.~\cite{Diaconescu01}), and
$R^\prime_T$ is the PV response function, defined as (again,
in the notation of Ref.~\cite{Diaconescu01})

\begin{equation}
R^\prime_T(q,\omega)=\overline{\sum_i}\sum_f\delta( \omega + m_i -E_f)\,
{\rm Im}\left[ {\bf j}_{fi}({\bf q}) \times {\bf j}^*_{fi}({\bf q}) \right]_z \ .
\end{equation}
Here the three-momentum transfer ${\bf q}$ has been taken to define
the spin-quantization axis (the $z$-axis), and
${\bf j}_{fi}({\bf q})\equiv \langle f|{\bf j}({\bf q})|i\rangle$ are
the matrix elements of the electromagnetic current.  In terms
of electric and magnetic multipole operators, the cross product
above is expressed as

\begin{equation}
{\rm Im}\left[ {\bf j}_{fi}({\bf q}) \times {\bf j}^*_{fi}({\bf q}) \right]_z \propto
\sum_{l \geq 1} {\rm Re}\Big[ \langle J_f||M_l(q)||J_i\rangle  
                              \langle J_f||E_l(q)||J_i\rangle^* \Big] \ ,
\end{equation}
and therefore vanishes unless (i) the initial and/or final states
do not have definite parity (as is the case here because of the presence
of PV $N$$N$ interactions) and/or (ii) the electric and magnetic
multipole operators have unnatural parities $(-)^{l+1}$ and
$(-)^l$, respectively, because of PV electromagnetic currents,
such as the one-body anapole current~\cite{anapole} and the
two-body currents due to PV $N$$N$ interactions~\cite{Carlson02}.
 
The one-body anapole current is taken as~\cite{anapole}

\begin{equation}
{\bf j}^{(1)}_{PV}({\bf q})= \frac{Q^2}{2\, m^2} \sum_i
\left[ a_S(Q^2) + a_V(Q^2) \tau_{i,z} \right] \bbox{\sigma}_i 
{\rm e}^{ {\rm i} {\bf q} \cdot {\bf r}_i }  \ ,
\end{equation}
where $Q^2$ is the squared four-momentum transfer ($Q^2$=$q^2-\omega^2$),
and the isoscalar and isovector anapole form factors are normalized as

\begin{equation}
a_{S,V}(0)=\frac{h_\pi \, g_\pi}{ 4 \sqrt{2} \pi^2 } \alpha_{S,V} \ ,
\end{equation}
with $\alpha_S$=1.6 and $\alpha_V$=0.4 from a calculation of pion-loop
contributions~\cite{anapole}.  Here $h_\pi$ and $g_\pi$ are weak and
strong $\pi$$N$$N$ coupling constants, respectively.  More recent estimates
of the nucleon anapole form factors give somewhat smaller values for
$\alpha_{S,V}$~\cite{Musolf95,Riska00,vanKolck00}. A complete treatment
would require estimates of short-distance contributions~\cite{Zhu00}
and electroweak radiative corrections, we postpone that discussion
to a later paper.\cite{Carlson02}

Only the PV two-body current associated with the $\pi$-exchange term
in the DDH interaction is included in the present calculations
(in addition, of course, to the PC two-body currents 
discussed in the previous section).  The PV currents from $\rho$
and $\omega$ exchanges have not yet been considered~\cite{Carlson02,Haxton02},
but are not expected to play a significant role.
One should note that at the higher momentum transfers
of interest here, 100--300 MeV/c, relevant for the SAMPLE
experiments, it is not possible to include the currents
through the Siegert theorem, they must be calculated explicitly.   

The calculation proceeds as discussed in Ref.~\cite{Diaconescu01}.
We have used the AV18 or Bonn-CD models (and associated currents) in
combination with the full DDH interaction (with coupling and cutoff
values as given in the previous section).  The final state, labeled
by the relative momentum$ {\bf p}$, pair spin and $z$-projection $S M_S$,
and pair isospin $T$ ($M_T$=0), is expanded in partial waves; PC
and PV interaction effects are retained in all partial waves
with $J \leq 5$, while spherical Bessel functions are employed
for $J>5$.  In the quasi-elastic regime of interest here, it has
been found that interaction effects are negligible for $J>5$.
 
In Fig.~1 we show the asymmetry for one of the SAMPLE kinematics for the AV18
plus full DDH interaction.  The total asymmetry is given by the sum of
the three contributions shown.  The contribution labeled \lq\lq DDH interaction\rq\rq~is
that originating directly from the DDH interaction, while that labeled
\lq\lq PV $\gamma$-couplings\rq\rq~is from the (one-body) anapole term and
the two-body pion current.  Results at lower $Q^2$ are similar.  The Bonn-CD
model leads to predictions, for both the asymmetry and inclusive
cross section, which are very close to those obtained with the AV18~\cite{Carlson02}.

In Fig.~2 we plot the asymmetry at the top of the quasi-elastic
peak as a function of the four-momentum transfer.  SAMPLE kinematics
have $Q^2$ equal to 0.1 (GeV/c)$^2$ and 0.043 (GeV/c)$^2$, respectively.
The leading $\gamma$-$Z$ term decreases in magnitude with $Q^2$, as expected.
This term does not contribute, of course, for the real photons
produced in the $\vec{n}p$ radiative capture.  Also shown in the plot is the
asymmetry obtained by retaining only the pion term in the DDH interaction,
with a coupling given by the \lq\lq best guess\rq\rq.

These results demonstrate that, in the kinematics of the SAMPLE
experiments, the asymmetry from the $\gamma$-$Z$ coupling
is two-orders of magnitude larger than that associated with the
PV hadronic weak interaction.  Hence even the largest estimates of the
weak $\pi$$N$$N$ coupling constant will not affect extractions
of nucleon matrix elements.

\section{Conclusion}

Parity-violating asymmetries in $np$ radiative capture and deuteron
electro-disintegration have been investigated within the framework
of the DDH model of the parity-violating hadronic weak interaction.
We find that the model-dependence of the $n$$p$-capture asymmetry
is quite small, at a level similar to the expected contributions
of the short-range parts of the interaction.  This process is in fact
dominated by the longer-range pieces of the interaction associated
with pion-exchange, and hence this experiment is a clean probe of
that physics.

Similarly, we find that the SAMPLE experiment measuring the asymmetry
in electron scattering on the deuteron is a very clean probe of nucleon
properties.  The processes associated with two-nucleons, including parity
violation in the deuteron and scattering wave functions, and the currents
associated with these interactions, play a very small role at reasonable
values of the momentum transfer.  These two experiments, then, probe 
distinct aspects of weak-interaction physics.

We like to thank C.-P.\ Liu, G.\ Pr\'ezeau, and M.J.\ Ramsey-Musolf
for making available to us the results of their calculation of
the asymmetry in quasi-elastic deuteron electro-disintegration prior to publication.
The work of J.C. and M.P. was supported by the U.S. Department of Energy
under contract W-7405-ENG-36, while that of R.S. was supported by the
U.S. Department of Energy contract DE-AC05-84ER40150 under which the
Southeastern Universities Research Association (SURA) operates the Thomas
Jefferson National Accelerator Facility.

%
%
%
%
\begin{table}[bthp]
\caption{Total $np$ capture cross-sections (in mb) and $\vec{n}p$ radiative capture 
asymmetries (in units of $10^{-8}$) in various models. Asymmetries are reported  for pion-exchange only and full DDH interactions.}
\begin{tabular}{|c|c|c|c|c|}
\hline
&\multicolumn{2}{c|}{Cross Section} & \multicolumn{2}{c|}{Asymmetry} \\
Interaction &Impulse Curr&Full Curr& pion-only & full DDH \\
\hline
AV18        & 304.6            & 334.2 & --4.98 & --4.85 \\
NIJM-I      & 305.4            & 332.5 & --5.11 & --4.95 \\ 
Bonn-CD     & 306.5            & 331.6 & --4.97 & --4.83 \\
\hline
\end{tabular}
\end{table}
%
%
%
%
\begin{table}[bthp]
\caption{Cumulative contributions to the $\vec{n}p$ radiative capture asymmetry
(in units of 10$^{-8}$) at thermal neutron energies
for the AV18 interaction and pion-exchange-only DDH interaction.
See text for explanation.}
\begin{tabular}{|c|c|c|}
\hline
          & AV18 (PC) Currents & DDH (PV) Currents \\
\hline
Impulse                          & --15.3 &      \\
$+\pi$                           & --48.3 & 44.2 \\
$+\rho$                          & --40.4 & 44.0 \\
$+p$-dependent                   & --43.8 & 44.0 \\
\hline
\end{tabular}
\end{table}
\begin{table}[bthp]
\caption{Cumulative contributions to the $\vec{n}p$ radiative capture asymmetry
(in units of 10$^{-8}$) at thermal neutron energies
for the AV6 interaction and pion-exchange-only DDH interaction.
Also listed is the asymmetry obtained with the Siegert form of 
the $E_1$ operator.  See text for explanation.}
\begin{tabular}{|c|c|c|c|}
\hline
          & AV6 (PC) Currents & DDH (PV) Currents & Total \\
\hline
Impulse                          & --18.5 &      & --18.5 \\
$+\pi$                            & --66.0 & 50.6 & --15.4 \\
$+\rho$                           & --56.7 & 50.4 & --6.32 \\
\hline
Siegert $E_1$                    &        &      & --6.15 \\
\hline
\end{tabular}
\end{table}
%
%
%
%
\begin{figure}[tbhp]
\resizebox{0.45\textwidth}{!}{\includegraphics{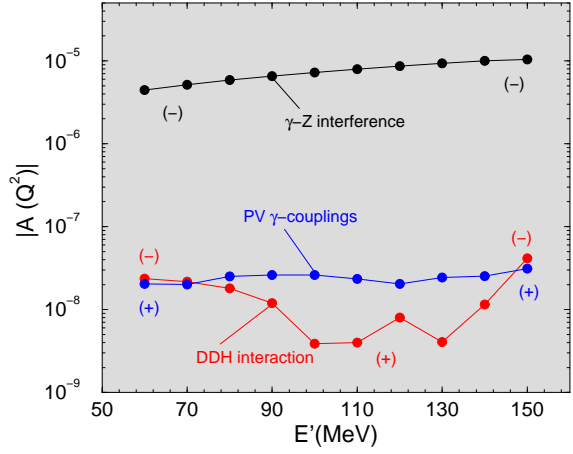}}
\caption{Contributions (in magnitude) to the longitudinal asymmetry
in the electro-disintegration of the deuteron as function of the final
electron energy.  The kinematical setting corresponds to one of those relevant
for the SAMPLE experiments: the initial electron energy is 193 MeV and
the electron scattering angle is 145.9$^\circ$.  The $(-)$ and $(+)$
symbols denote the signs of the various contributions.  See text for
further explanations.}
\label{fig:asy}
\end{figure}
\begin{figure}[tbh]
\resizebox{0.45\textwidth}{!}{\includegraphics{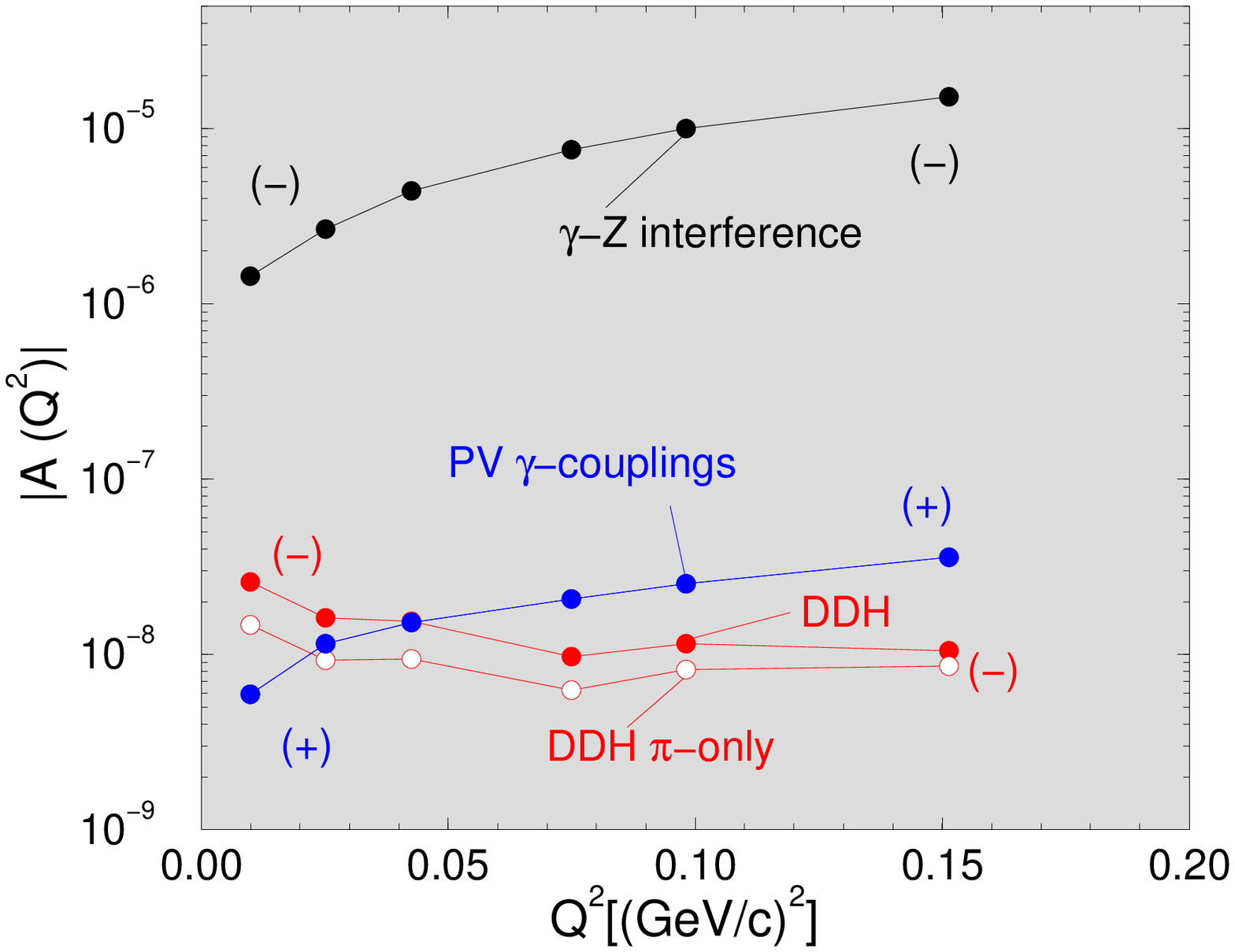}}
\caption{Contributions (in magnitude) to the longitudinal asymmetry
in the deuteron electro-disintegration at the top of the quasi-elastic
peak, plotted as function of the four-momentum transfer $Q^2$.  The $(-)$
and $(+)$ symbols denote the signs of the various contributions.  See text for
further explanations.}
\label{fig:asyq2}
\end{figure}

\begin{thebibliography}{10}
%
%
\bibitem{Berdoz01} A.R.\ Berdoz {\it et al.},
                 Phys.\ Rev.\ Lett.\ {\bf 87}, 272301 (2001).
%
\bibitem{npexpt} W.M. Snow {\it et al.}, Nucl. Inst. and
Meth. A {\bf 440}  729 (2000).
%
\bibitem{ddhpv} B.\ Desplanques, J.F.\ Donoghue, and B.R.\ Holstein,
                Ann.\ Phys.\ (N.Y.) {\bf 124}, 449 (1980).
%
\bibitem{Hasty00} R.\ Hasty {\it et al.},
                  Science {\bf 290}, 2117 (2000).
%
\bibitem{samplereview} D.H.\ Beck and R.D.\ McKeown,
                       Ann.\ Rev.\ Nucl.\ Part.\ Sci.\ {\bf 51}, 189 (2001).
%
\bibitem{Carlson02} J.\ Carlson, M.\ Paris, and R.\ Schiavilla,
                    in preparation.
%
\bibitem{npdgtheory1} B.\ Desplanques,
                      Nucl.\ Phys.\ {\bf A242}, 423 (1975); {\it ibidem} {\bf A335}, 147 (1980).
%
\bibitem{npdgtheory2} B.H.J.\ McKellar,
                      Nucl.\ Phys.\ {\bf A254}, 147 (1975).
%
\bibitem{flourine} E.G.\ Adelberger and W.C.\ Haxton,
                   Ann.\ Rev.\ Nucl.\ Part.\ Sci.\ {\bf 35}, 501 (1985).
%
\bibitem{cesium} V.V.\ Flambaum and D.W.\ Murray,
                 Phys.\ Rev.\ C {\bf 56}, 1641 (1997). 
%
\bibitem{haxtonreview} W.C.\ Haxton and C.E.\ Wieman,
                       Ann.\ Rev.\ Nuc.\ Part.\ Sci.\ {\bf 51}, 261 (2001).
%
\bibitem{Haxton02} W.C.\ Haxton, C.-P. Liu, and M.J.\ Ramsey-Musolf,
                  Phys.\ Rev.\ C {\bf 65}, 045502 (2002).
%
\bibitem{bowmanprivate} D.\ Bowman, private communication.
%
\bibitem{pppvcalc} J.\ Carlson, R.\ Schiavilla, V.R.\ Brown, and B.F.\ Gibson,
                   Phys.\ Rev.\ C {\bf 63}, 024003 (2002).
%
\bibitem{Pirner73} H.J.\ Pirner and D.O.\ Riska,
                   Phys.\ Lett.\ {\bf B44}, 151 (1973).
%
\bibitem{av18} R.B.\ Wiringa, V.G.J.\ Stoks, and R.\ Schiavilla,
               Phys.\ Rev.\ C {\bf 51}, 38 (1995).
%
\bibitem{bonncd} R.\ Machleidt,
                 Phys.\ Rev.\ C {\bf 63}, 024001 (2001).
%
\bibitem{nijm1} V.G.J.\ Stoks, R.A.M.\ Klomp, C.P.F.\ Terheggen, and J.J.\ de Swart,
                Phys.\ Rev.\ C {\bf 49}, 2950 (1994).
%
\bibitem{pccurrents} J.\ Carlson and R.\ Schiavilla,
                     Rev.\ Mod.\ Phys.\ {\bf 70}, 743 (1998). 
%
\bibitem{npxsec1} A.E.\ Cox, S.A.R.\ Wynchank, and C.H.\ Collie,
                  Nucl.\ Phys.\ {\bf 74}, 497 (1965).
%
\bibitem{npxsec2} S.F.\ Mughabghab, M.\ Divadeenam, and N.E.\ Holden, 
{\it Neutron Cross Sections from Neutron Resonance Parameters and Thermal Cross Sections}
(Academic Press, London, 1981), http://isotopes.lbl.gov/ngdata/sig.htm.
%
\bibitem{Desplanques01} B.\ Desplanques,
                        Phys.\ Lett.\ {\bf B512}, 305 (2001).
%
\bibitem{Viviani00} M.\ Viviani, A.\ Kievsky, L.E.\ Marcucci, S.\ Rosati, and R.\ Schiavilla,
                    Phys.\ Rev.\ C {\bf 61}, 064001 (2000).
%
\bibitem{Schiavilla89} R.\ Schiavilla, V.R.\ Pandharipande, and D.O.\ Riska,
                       Phys.\ Rev.\ C {\bf 40}, 2294 (1989).
%
\bibitem{Tsushima93} K.\ Tsushima, D.O.\ Riska, and P.G.\ Blunden,
                     Nucl.\ Phys.\ {\bf A559}, 543 (1993).
%
\bibitem{Wiringa02} R.B.\ Wiringa and S.C.\ Pieper, 
                    Phys.\ Rev.\ Lett.\ {\bf 89}, 182501 (2002).
%
\bibitem{Diaconescu01} L.\ Diaconescu, R.\ Schiavilla, and U.\ van Kolck,
                       Phys.\ Rev.\ C {\bf 63}, 044007 (2001). 
%
\bibitem{Hwang80} W.-Y. P.\ Hwang and E.M.\ Henley,
                  Ann.\ Phys.\ (N.Y.) {\bf 129}, 47 (1980).
%
\bibitem{Hwang81} W.-Y. P.\ Hwang, E.M.\ Henley, and G.A.\ Miller,
                  Ann.\ Phys.\ (N.Y.) {\bf 137}, 378 (1981).
%
\bibitem{anapole} W.C.\ Haxton, E.M.\ Henley, and M.J.\ Musolf,
                  Phys.\ Rev.\ Lett.\ {\bf 63}, 949 (1989).
%
\bibitem{Musolf95} M.J.\ Musolf and B.R.\ Holstein,
                   Phys.\ Rev.\ D {\bf 43}, 2956 (1991).
%
\bibitem{Riska00} D.O.\ Riska,
                  Nucl.\ Phys.\ A {\bf 678}, 79 (2000).
%
\bibitem{vanKolck00} C.M.\ Maekawa and U.\ van Kolck,
                     Phys.\ Lett.\ {\bf B478}, 73 (2000).
%
\bibitem{Zhu00} S.L. Zhu {\it et al.}, Phys. Rev. D{\bf62}, 033008 (2000).
%
%
%
\end{thebibliography}
\end{document}